\documentclass[prd,nofootinbib,showpacs,preprintnumbers,amsmath,amssymb]{revtex4}
\usepackage{graphicx}% Include figure files
\usepackage{dcolumn}% Align table columns on decimal point
\usepackage{bm}% bold math

\def\bear{\begin{eqnarray}}
\def\ear{\end{eqnarray}}

\begin{document}

\title{Massive Scalar Field Quantum Cosmology}

\author{Sang Pyo Kim}\email{sangkim@kunsan.ac.kr}
\affiliation{Department of Physics, Kunsan National University, Kunsan 573-701, Korea}

\medskip

\date{\today}

\begin{abstract}
We study the third quantized formulation of the massive scalar field quantum cosmology for the Friedmann-Robertson-Walker universe. The Hamiltonian is equivalent to an infinite number of coupled oscillators whose couplings and frequencies are intrinsic time-dependent. We propose the invariant operators whose eigenstates provide the exact wave functions of the universe.
\end{abstract}
\pacs{98.80.Qc, 04.60.-m, 04.60.Kz, 98.80.Cq}

\maketitle

\section{Introduction}

In quantum cosmology both spacetime geometries and matter fields are quantized and obey the Wheeler-DeWitt (WDW) equation. Minisuperspace quantum cosmological models \cite{hawking,hartle-hawking} have a few degrees of freedom and hence allow one to find wave functions exactly or approximately, though the solution space of the WDW equation is infinite dimensional. The geometry of the Friedmann-Robertson-Walker (FRW) universe is prescribed by a single scale factor and is thus the simplest minisuperspace cosmological model. The homogeneous and isotropic universe of FRW geometry together with a homogeneous scalar field is still favored by the current observational data. In fact, the inflationary cosmology with a large-field inflaton with power law less than two is consistent with nine-year WMAP data \cite{9WMAP} and Planck 2013 results \cite{Planck}.

The massive scalar field quantum cosmology (MSQC), which naturally includes fluctuations of geometry, may be a viable inflationary model.
The MSQC has only two degrees of freedom: the scale factor of the FRW geometry and the homogeneous scalar field. However, MSQC is not just a quantum mechanical system since it is governed by a relativistic wave equation in the minisuperspace of the three-geometry and the scalar field. Furthermore, the scalar field is parametrically coupled to the scale factor in the WDW equation, which is analogous to the Klein-Gordon equation with a time-dependent mass.

The parametric interaction of the massive scalar field with the scale factor prevents the WDW equation from being simply separated by the harmonic wave functions for the scalar field. Provided that the scale factor is used as an intrinsic time, the WDW equation has the Cauchy initial value problem and is equal to a time-dependent matrix equation, whose solutions show couplings among harmonic wave functions for the massive scalar field during the evolution \cite{kim-page92,kim92}. The Feshbach-Villars equation \cite{feshbach-villars} allows one to express the WDW equation in the first order formalism, which has the same form as the Schr\"{o}dinger equation with a non-Hermitian Hamiltonian \cite{mostafazadeh98,mostafazadeh04}, which does not, however, possess any quantum invariant that has been found for time-dependent oscillators by Lewis and Riesenfeld \cite{lewis-riesenfeld}.

In this paper we show that MSQC in the third quantized formulation is equivalent to an infinite number of coupled oscillators with intrinsic time-dependent frequencies. We further advance quantum invariants for MSQC in the third quantized formulation that play the role of time-dependent annihilation and creation operators in constructing all quantum states, which generalize those for decoupled time-dependent oscillators \cite{MMT70,kim-lee00,kim-page01}.

The organization of this paper is as follows. In Sec. II we review the WDW equation for MSQC in the first order formalism and show that the WDW equation does not have any quantum invariant in the first order formalism. In Sec. III we formulate the WDW equation in the third quantization and express the Hamiltonian in the third quantization as an infinite number of coupled time-dependent oscillators. We advance the invariant operators and discuss their physical implications.

\section{First Order Formalism for WDW Equation}

The WDW equation for the FRW universe with a massive scalar field is [in units of $c=\hbar=G=1$]
\begin{eqnarray}
\Bigl[ \frac{\partial^2}{\partial \alpha^2} + \hat{H}_m (\phi, \alpha) \Bigr] \Psi (\alpha, \phi) = 0, \label{wdw eq}
\end{eqnarray}
where $e^{\alpha}$ is the scale factor and
\begin{eqnarray}
\hat{H}_m (\alpha) = - \frac{\partial^2}{\partial \phi^2} + m^2 e^{6 \alpha} \phi^2 - k e^{4 \alpha}, \label{mass ham}
\end{eqnarray}
is the Hamiltonian for the massive scalar field.
Here the last term comes from the scalar curvature of the three-geometry with $k = 1, 0$ and $-1$ for a closed, spatially flat and open universe, respectively. The Hamiltonian (\ref{mass ham}) describes a harmonic oscillator and has an  $su(1,1)$ algebra \cite{wybourne}.

The wave functions for Eq. (\ref{wdw eq}) have not been known in spite of intensive investigations. The mathematical difficulty lies in the parametric dependence of the Hamiltonian (\ref{mass ham}), in which the frequency depends on the intrinsic time $\alpha$. It is worthy to mention that scalar quantum electrodynamics in the vector potential $\vec{A}(t) = \vec{B}(t) \times \vec{x}/2$ has a similar structure as the WDW equation \cite{kim13}. In contrast to parametrically interacting relativistic equations, time-dependent harmonic oscillators have a quadratic invariant found by Lewis and Riesenfeld, whose eigenstates provide exact quantum states up to time-dependent phase factors \cite{lewis-riesenfeld}.

Hence it is interesting to investigate whether the WDW equation in the first order formalism possesses such a quantum invariant or not. Employing the Feshbach-Villars equation for a scalar field \cite{feshbach-villars}, Mostafazadeh expressed Eq. (\ref{wdw eq}) in the first order formalism \cite{mostafazadeh98,mostafazadeh04}
\begin{eqnarray}
i \frac{\partial }{\partial \alpha}
\begin{pmatrix}
  \Psi + i \frac{\partial \Psi}{\partial \alpha} \\
\Psi - i \frac{\partial \Psi}{\partial \alpha} \end{pmatrix}
= \frac{i}{2} \begin{pmatrix}
  1 + \hat{H}_m & -(1 - \hat{H}_m) \\
  1- \hat{H}_m  & -(1+ \hat{H}_m) \end{pmatrix} \begin{pmatrix}
  \Psi + i \frac{\partial \Psi}{\partial \alpha} \\
\Psi - i \frac{\partial \Psi}{\partial \alpha} \end{pmatrix}. \label{most eq}
\end{eqnarray}
Though Eq. (\ref{most eq}) may be interpreted as the Schr\"{o}dinger equation with the non-Hermitian Hamiltonian
\begin{eqnarray}
{\bf H} =  \frac{i}{2} \begin{pmatrix}
  1 + \hat{H}_m & -(1 - \hat{H}_m) \\
  1- \hat{H}_m  & -(1+ \hat{H}_m) \end{pmatrix},
\end{eqnarray}
there does not exist any invariant in the algebra of $su(2) \otimes su(1,1)$, which satisfies
\begin{eqnarray}
i \frac{\partial {\bf I}}{\partial \alpha} + [{\bf I}, {\bf H}] = 0. \label{1st inv}
\end{eqnarray}

On the other hand, Kim introduced another form of the first order formalism \cite{kim-page92,kim92}
\begin{eqnarray}
\frac{\partial }{\partial \alpha}
\begin{pmatrix}
  \Psi \\
\frac{\partial \Psi}{\partial \alpha} \end{pmatrix}
= \begin{pmatrix}
  0 & 1 \\
 - \hat{H}_m  & 0 \end{pmatrix} \begin{pmatrix}
  \Psi \\
\frac{\partial \Psi}{\partial \alpha} \end{pmatrix}. \label{kim eq}
\end{eqnarray}
A stratagem is to use the eigenstates of the massive scalar field
\begin{eqnarray}
\hat{H}_m \vert \Phi_n (\phi; \alpha) \rangle = m e^{3 \alpha} (2n + 1) \vert \Phi_n (\phi; \alpha) \rangle \label{num st}
\end{eqnarray}
and to expand the wave function in the form
\begin{eqnarray}
\vert \Psi (\alpha, \phi) \rangle = \vert \vec{\Phi} (\phi; \alpha) \rangle^{T} \cdot \vec{\psi} (\alpha), \label{wav ex}
\end{eqnarray}
where $\vert \vec{\Phi} (\phi; \alpha) \rangle$ is a column vector of eigenstates and $T$ denotes the transpose. Then, the wave function in the first order formalism can be solved in terms of the scattering matrix \cite{kim92}
\begin{eqnarray}
\begin{pmatrix}
  \vert \Psi (\alpha, \phi) \rangle \\
\frac{\partial}{\partial \alpha} \vert \Psi (\alpha, \phi) \rangle \end{pmatrix}
= \begin{pmatrix}
 \vert \vec{\Phi} (\phi; \alpha) \rangle^{T}  & 0 \\
 0 & \vert \vec{\Phi} (\phi; \alpha) \rangle^{T} \end{pmatrix} {\cal T}
 \exp \Biggl[ \int_{\alpha_0}^{\alpha} \begin{pmatrix}
  \Omega  & I \\
 - E (\alpha) & \Omega \end{pmatrix} d \alpha \Biggr] \begin{pmatrix}
\vec{\psi} (\alpha_0)\\
\frac{d  \vec{\psi} (\alpha_0)}{d \alpha} \end{pmatrix}.
\end{eqnarray}
Here the Cauchy initial data are
\begin{eqnarray}
\vec{\psi} (\alpha_0) &=& \int d\phi \langle \vec{\Phi} (\phi; \alpha_0) \vert \Psi (\alpha_0, \phi) \rangle, \nonumber\\
\frac{d  \vec{\psi} (\alpha_0)}{d \alpha} &=& \int d\phi \langle \vec{\Phi} (\phi; \alpha_0) \vert \frac{\partial}{ \partial \alpha}
\vert \Psi (\alpha_0, \phi) \rangle,
\end{eqnarray}
and ${\cal T}$ denotes the time-ordered integral and in the number-state representation (\ref{num st})
the coupling matrix and the energy-eigenvalue matrix are given by
\begin{eqnarray}
\Omega = \frac{3}{4} (\hat{a}^2 - \hat{a}^{+2} ), \quad E (\alpha) = m e^{3 \alpha} ( 2\hat{a}^{+} \hat{a} +
1) - k e^{4 \alpha}. \label{c-matrix}
\end{eqnarray}
The off-diagonal matrix $\Omega$ causes continuous transitions among different number states of the massive scalar field. Furthermore, in the limit of
 $\alpha$, $\alpha_0 = - \infty$ with a finite difference $\Delta \alpha= \alpha - \alpha_0$,
$E(\alpha)$ is exponentially suppressed, so the scattering matrix simply becomes
\begin{eqnarray}
{\cal T}
 \exp \Biggl[ \int_{\alpha_0}^{\alpha} \begin{pmatrix}
  \Omega  & I \\
 - E (\alpha) & \Omega \end{pmatrix} d \alpha \Biggr] =
 \exp \Biggl[ \begin{pmatrix}
  \Omega  & I \\
 0 & \Omega \end{pmatrix} (\alpha - \alpha_0) \Biggr]
 = \begin{pmatrix}
 e^{\Omega \Delta \alpha}   & \Delta \alpha e^{\Omega \Delta \alpha} \\
 0 & e^{\Omega \Delta \alpha} \end{pmatrix}.
\end{eqnarray}
Therefore, the wave function near the big bang singularity infinitely oscillates as
\begin{eqnarray}
\vert \Psi (\alpha, \phi) \rangle = \vert \vec{\Phi} (\alpha) \rangle^{T}  e^{\Omega \Delta \alpha}
\Bigl[ \vec{\psi} (\alpha_0)+ \Delta \alpha \frac{\partial \vec{\psi}(\alpha_0)}{\partial \alpha}
\Bigr].
\end{eqnarray}
However, the magnitude of the wave function
\begin{eqnarray}
\langle \Psi (\alpha, \phi) \vert \Psi (\alpha, \phi) \rangle = \Bigl| \vec{\psi} (\alpha_0)+
\Delta \alpha \frac{\partial \vec{\psi} (\alpha_0)}{\partial \alpha} \Bigr|^2
\end{eqnarray}
is determined only by the Cauchy data, independently of the evolution of the scalar field and the geometry, which was first observed
in Ref. \cite{kim92}. This implies that any regular wave function should have this scale invariance near the big bang singularity.

\section{Third Quantized Formulation}

The WDW equation (\ref{wdw eq}) for the FRW universe is a hyperbolic equation, in particular, when $\alpha$ is regarded as a time-like variable in the minisuperspace of the metric and the field. In general, the superspace for a globally hyperbolic spacetime with or without matter fields has a Lorentzian signature. The WDW equation is thus analogous to a Klein-Gordon equation, in which one of the superspace variables plays the role of the intrinsic time. It should be noted, however, that the time cannot be uniquely determined: any functional of the chosen time-like variable also becomes another time-like variable. But the hyperbolic nature of the WDW equation does not depend on the redefintion of the time-like variable. The same conceptual problem occurs in the foliation of a hyperbolic spacetime. In this paper we adopt the picture that $\alpha$ is an intrinsic time in the WDW equation for MSQC.

In the third quantized formulation \cite{mcguigan88,giddings-strominger89,mcguigan89,hosoya-morikawa89,peleg91},
the WDW equation for MSQC derives from the action in the minisuperspace
\begin{eqnarray}
{\cal S} = \frac{1}{2} \int d \alpha d \phi \Bigl[ \Bigl(\frac{\partial \Psi}{\partial \alpha} \Bigr)^2
- \Bigl(\frac{\partial \Psi}{\partial \phi} \Bigr)^2 - V (\phi, \alpha) \Psi^2\Bigr], \label{3-act}
\end{eqnarray}
where
\begin{eqnarray}
V (\phi, \alpha) = m^2 e^{6 \alpha} \phi^2 - k e^{4 \alpha}.
\end{eqnarray}
The variation $\delta {\cal S}/\delta \Psi$ leads to the WDW equation (\ref{wdw eq}). We shall focus on the massive scalar field and consider the massless scalar field as the limiting case of $m=0$.

To find the Hamiltonian from the action (\ref{3-act}), we expand the wave function according to Eq. (\ref{wav ex}) to get
\begin{eqnarray}
\int d \phi \Psi^2 = \vec{\psi}^T \cdot \vec{\psi},
\end{eqnarray}
and
\begin{eqnarray}
\int d \phi \Bigl(\frac{\partial \Psi}{\partial \alpha} \Bigr)^2 = (\dot{\vec{\psi}}^T + \vec{\psi}^T \Omega^T) \cdot
(\dot{\vec{\psi}} + \Omega \vec{\psi}),
\end{eqnarray}
where $\Omega$ is the coupling matrix (\ref{c-matrix}). Hence the Lagrangian is given by
\begin{eqnarray}
{\cal L} = \frac{1}{2} (\dot{\vec{\psi}}^T + \vec{\psi}^T \Omega^T) \cdot
(\dot{\vec{\psi}} + \Omega \vec{\psi}) - \frac{1}{2} \vec{\psi}^T E (\alpha) \vec{\psi}
\end{eqnarray}
Introducing the canonical momentum vector
\begin{eqnarray}
\vec{\pi} = \frac{\partial {\cal L}}{\partial \dot{\vec{\psi}}^T} = \dot{\vec{\psi}} + \Omega \vec{\psi}
\end{eqnarray}
and using $\Omega^T = - \Omega$, we obtain the Hamiltonian
\begin{eqnarray}
{\cal H} (\alpha) = \frac{1}{2} \vec{\pi}^T \cdot \vec{\pi} - \vec{\pi}^T \Omega \vec{\psi} +
\frac{1}{2} \vec{\psi}^T E  \vec{\psi}. \label{3-ham}
\end{eqnarray}
The quantum law for the universe is the functional Schr\"{o}dinger equation
\begin{eqnarray}
i \frac{\partial}{\partial \alpha} \Psi (\alpha) = \hat{\cal H} (\alpha) \Psi (\alpha).
\end{eqnarray}

The Hamiltonian (\ref{3-ham}) is an infinite system of coupled oscillators with time-dependent frequencies $E(\alpha)$.
We propose a pair of invariant operators of the form
\begin{eqnarray}
\vec{\cal A} (\alpha) &=& i [{\cal O} \vec{\pi} - (\dot{\cal O} + {\cal O} \Omega^T) \vec{\psi}], \nonumber\\
\vec{\cal A}^{\dagger} (\alpha) &=& - i [\vec{\pi}^{\dagger} {\cal O}^{\dagger}  - \vec{\psi}^{\dagger} (\dot{\cal O}^{\dagger} + \Omega^* {\cal O}^{\dagger})],
\end{eqnarray}
which satisfy the Liouville-von Neumann equation
\begin{eqnarray}
i \frac{\partial}{\partial \alpha} \begin{pmatrix}
  \vec{\cal A} (\alpha) \\
 \vec{\cal A}^{\dagger} (\alpha) \end{pmatrix} + \Biggl[\begin{pmatrix}
  \vec{\cal A} (\alpha) \\
 \vec{\cal A}^{\dagger} (\alpha) \end{pmatrix}, \hat{\cal H} (\alpha) \Biggr] = 0.
\end{eqnarray}
Here ${\cal O}$ is a matrix satisfying
\begin{eqnarray}
\ddot{\cal O} + 2 \dot{\cal O} \Omega^T + {\cal O} (E + \dot{\Omega}^T + (\Omega^T)^2) = 0, \label{O eq}
\end{eqnarray}
and is chosen such that $\vec{\cal A}$ and $\vec{\cal A}^{\dagger}$ satisfy the equal-time commutator
\begin{eqnarray}
[\vec{\cal A} (\alpha), \vec{\cal A}^{\dagger} (\alpha)] = I.
\end{eqnarray}
It is interesting to compare Eq. (\ref{O eq}) with the WDW equation in the vector notation
\begin{eqnarray}
\ddot{\vec{\psi}} - 2 \Omega \dot{\vec{\psi}} + (E - \dot{\Omega} + (\Omega)^2) \vec{\psi} = 0.
\end{eqnarray}

In the special case of a massless scalar field ($m=0$), we take the limit $\Omega=0$, which has been elaborated in Ref. \cite{kim12}.
The Hamiltonian, after decomposing by the Fourier modes $\vert \Phi \rangle = e^{ip \phi}$,
is another infinite system of decoupled time-dependent oscillators, so the wave function is the product of wave functions for Fourier cosine and sine modes:
\begin{eqnarray}
\Psi (\alpha, \phi) = \prod_{(\pm) p} \Psi_{(\pm) p} (\alpha) e^{i p \phi},
\end{eqnarray}
where
\begin{eqnarray}
i \frac{\partial}{\partial \alpha} \Psi_{(\pm) p} (\alpha) = \hat{\cal H}_{(\pm) p} (\alpha) \Psi_{(\pm) p} (\alpha). \label{massless osc}
\end{eqnarray}
The invariant operators $\vec{\cal A}$ and $\vec{\cal A}^{\dagger}$ decouple among themselves and are given by
\begin{eqnarray}
\hat{\cal A}_{(\pm) p} (\alpha) &=& i [u^*_{(\pm) p} \hat{\pi}_{(\pm) p} - \dot{u}^*_{(\pm) p} \hat{\psi}_{(\pm) p} ], \nonumber\\
\hat{\cal A}^{\dagger}_{(\pm) p} (\alpha) &=& -i [u_{(\pm) p} \hat{\pi}_{(\pm) p} - \dot{u}_{(\pm) p} \hat{\psi}_{(\pm) p} ], \label{massless pair}
\end{eqnarray}
where $u_{(\pm) p}$ satisfies the mode equation
\begin{eqnarray}
\ddot{u}_{(\pm) p} + (p^2 - k e^{4 \alpha}) u_{(\pm) p} =0, \label{massless cl eq}
\end{eqnarray}
and the Wronskian condition ${\rm Wr} [u_{(\pm) p}, u^*_{(\pm) p}] = i$.
The physical implications of the third quantized formulation have been discussed in detail in Ref. \cite{kim12}.

\section{Conclusion}

In this paper we have investigated the massive scalar field quantum cosmology in the view of nine-year WMAP data \cite{9WMAP} and Planck 2013 results \cite{Planck}. Though the massive scalar field alone seems to be likely excluded, the curvature-square model by Starobinsky turns out a viable scenario \cite{9WMAP,Planck}. As the curvature-square term originates from spacetime fluctuations probed by matter fields or gravity itself, it would be worthy to study quantum cosmology with a massive scalar field. In fact, the Wheeler-DeWitt equation necessarily involves quantum fluctuations of the geometries and the wave functions contain higher curvature effects. The de Broglie-Bohm interpretation of the wave function results in semiclassical quantum gravity with non-trivial higher curvature terms when the Planckian scale gravity separates from the massive scalar field via the Born-Oppenheimer idea \cite{kim95,kim97}.

The wave function for the massive scalar field quantum cosmology could be separated as a vector equation due to the continuous transitions among the eigenstates of the massive scalar field, which depends parametrically on the intrinsic time \cite{kim-page92,kim92}. This coupling matrix survives as an effective gauge potential in semiclassical quantum gravity \cite{kim95,kim97}. In the first order formalism the Wheeler-DeWitt equation could be analytically expressed by introducing the scattering matrix, which provides a perturbation method for wave functions. However, in the third quantized formulation the massive scalar field quantum cosmology could be expressed as an infinite system of coupled intrinsic time-dependent oscillators as shown in this paper. We have advanced the invariant operators for the resulted coupled oscillators, which will provide the exact wave functions of the universe. The quantum nature of the universe will be further studied from the view of the recent precision observation of CMB data in a future publication.

\acknowledgments
The author would like to thank Misao Sasaki and Takahiro Tanaka for the warm hospitality at Gravitation and Cosmology 2012 (GC2012), Yukawa Institute for Theoretical Physics, Kyoto University, where this paper was initiated. This work was supported in part by Basic Science Research Program through the National Research Foundation of Korea (NRF) funded by the Ministry of Education, Science and Technology (2012R1A1B3002852).


\begin{thebibliography}{99}

\bibitem{hawking} S.~W.~Hawking, ``The quantum state of the universe,'' Nucl.\ Phys.\ B {\bf 239}, 257 (1984).

\bibitem{hartle-hawking} J.~B.~Hartle and S.~W.~Hawking, ``Wave function of the Universe,'' Phys.\ Rev.\ D {\bf 28}, 2960 (1983).

\bibitem{9WMAP} G.~Hinshaw et al, ``Nine-Year Wilkinson Microwave AnisotropyPROBE (WMAP) Observations: Cosmological Parameter Results,'' [arXiv:1212.5226].

\bibitem{Planck} P.~A.~R.~Abe et al (Planck Collaboration), ``Planck 2013 results. XVI. Cosmological Parameters,'' [arXiv:1303.5076].

\bibitem{kim-page92} S.~P.~Kim and D.~N.~Page, ``Wormhole spectrum of a quantum Friedmann-Robertson-Walker cosmology minimally coupled to a power-law scalar field and the cosmological constant,'' Phys.\ Rev.\ D {\bf 45}, 45, R3296 (1992).

\bibitem{kim92} S.~P.~Kim, ``Quantum mechanics of conformally and minimally coupled Friedmann-Roberston-Walker cosmology,'' Phys.\ Rev.\ D {\bf 46}, 3403 (1992).

\bibitem{feshbach-villars} H.~Feshbach and F.~Villars, ``Elementary Relativistic Wave Mechanics of Spin 0 and Spin 1/2 Particles,'' Rev.\ Mod.\ Phys.\ {\bf 30}, 24 (1958)

\bibitem{mostafazadeh98} A.~Mostafazadeh,  ``Two-component formulation of the Wheeler–DeWitt equation,'' J.\ Math.\ Phys.\ {\bf 39}, 4499 (1998).

\bibitem{mostafazadeh04}  A.~Mostafazadeh,  ``Quantum mechanics of Klein–Gordon-type fields and quantum cosmology,'' Ann.\ Phys.\ {\bf 309}, 1 (2004).

\bibitem{lewis-riesenfeld} H.~R.~Lewis and W.~B.~Riesenfeld, ``An Exact Quantum Theory of the Time-Dependent Harmonic Oscillator
and of a Charged Particle in a Time‐Dependent Electromagnetic Field,'' J.\ Math.\ Phys.\ {\bf 10}, 1458 (1969).

\bibitem{MMT70} I.~A.~Malkin, V.~I.~Man'ko and D.~A.~Trifonov, ``Coherent States and Transition Probabilities in a Time-Dependent
Electromagnetic Field,'' Phys.\ Rev.\ D {\bf 2}, 1371 (1970).

\bibitem{kim-lee00} S.~P.~Kim and C.~H.~Lee, ``Nonequilibrium quantum dynamics of second order phase transitions,''
Phys.\ Rev.\ D {\bf 62}, 125020 (2000).

\bibitem{kim-page01} S.~P.~Kim and D.~N.~Page, ``Classical and quantum action-phase variables for time-dependent oscillators,''
Phys.\ Rev.\ A {\bf 64}, 012104 (2001).

\bibitem{wybourne} G.~B.~Wybourne, {\it Classical Groups for Physicists} (Wiley, New York, 1974).

\bibitem{kim13} S.~P.~Kim, ``Scalar QED in Time-Dependent Magnetic Field,'' in preparation.

\bibitem{giddings-strominger89} S.~B.~Giddings and A.~Strominger, Baby Universes, Third Quantization and The Cosmological Constant,
Nucl.\ Phys.\ B {\bf 321}, 481 (1989).

\bibitem{mcguigan88} M.~McGuigan, ``Third quantization and the Wheeler-deWitt equation,'' Phys.\ Rev.\ D {\bf 38},
3031 (1988).

\bibitem{mcguigan89} M.~McGuigan, ``Universe creation from the third-quantized vacuum,'' Phys.\ Rev.\ D {\bf 39},
2229 (1989).

\bibitem{hosoya-morikawa89} A.~Hosoya and M.~Morikawa, ``Quantum field theory of the Universe,'' Phys.\ Rev.\
D {\bf 39}, 1123 (1989).


\bibitem{peleg91} Y.~Peleg, ``On the third quantization of general relativity,'' Class.\ Quantum Grav.\ {\bf 8}, 827
(1991).

\bibitem{kim12} S.~P.~Kim, ``Third Quantization and Quantum Universes,'' CosPA 2012 Proceedings [arXiv:1212.5355].

\bibitem{kim95} S.~P.~Kim, ``New asymptotic expansion method for the Wheeler-DeWitt equation,'' Phys.\ Rev.\ D {\bf 52}, 3382 (1995).

\bibitem{kim97} S.~P.~Kim, ``Problem of unitarity and quantum corrections in semiclassical quantum gravity,''
Phys.\ Rev.\ D {\bf 55}, 7511 (1997).

\end{thebibliography}
\end{document}